\documentclass[ukenglish,notitlepage,a4paper,12pt]{article}
\usepackage{sw20elba}
\usepackage[dvips]{graphicx}

\begin{document}

\author{\textbf{George Jaroszkiewicz} \\
School of Mathematical Sciences,\\
University of Nottingham, UK}
\title{\textbf{The running of the Universe and}\\
\textbf{the quantum structure of time}}
\date{Wednesday 2$^{nd}$ May $2001$}
\maketitle

\begin{abstract}
Some principles underpinning the running of the Universe are discussed. The
most important, the machine principle, states that the Universe is a fully
autonomous, self-organizing and self-testing quantum automaton. Continuous
space and time, consciousness and the semi-classical observers of quantum
mechanics are all emergent phenomena not operating at the fundamental level
of the machine Universe. Quantum processes define the present, the interface
between the future and the past, giving a time ordering to the running of
the Universe which is non-integrable except on emergent scales. A
diagrammatic approach is used to discuss the quantum topology of the EPR
paradox, particle decays and scattering processes. A toy model of a
self-referential universe is given.
\end{abstract}

\section{Introduction}

In this paper some principles underpinning the running of the Universe on a
fundamental level are discussed. These are related to ideas about discrete
spacetime discussed recently by various authors [1-5] but important
differences exist. In particular, it is assumed here that the Universe runs
according to the following principle encapsulated by Bragg \cite
{BRAGG'S-PRINCIPLE}:

\begin{quote}
\textbf{Bragg's principle:} ``\emph{Everything in the future is a wave,
everything in the past is a particle''}.
\end{quote}

Whilst matters cannot be quite as simple as that, this principle says that
quantum processes define the fleeting moment of the present, which is the
transition from the unformed and uncertain future to a classically fixed and
unique past \cite{WHITROW:80}. In other words, time is a quantum phenomenon.

Bragg's principle reflects the human experience of time, the feeling that
the past and the future are neither equivalent nor symmetric about the
present, contrary to the temporal symmetry inherent in classical mechanics.
For example, Maxwell's equations have both advanced and retarded solutions,
requiring the former to be excluded by hand in order to retain classical
causality. Despite such examples and the fact that the standard formulation
of quantum mechanics makes a distinction between the past and the future,
state reduction (wave-function collapse) is commonly regarded as an ugly
idea best eliminated if possible from an otherwise elegant theory.

The problem comes from the concept of \emph{observer }in quantum
mechanics. Observers are free to prepare various states and then decide on
various experiments to do on them. This adds an unfortunate flavour to an
otherwise mathematically elegant theory because of its undertones of
subjectivity tacked onto an objective physics.

State reduction as the origin of time has been discussed before \cite
{OMNES:94} but in the context of continuous time. In this paper the
spacetime continuum is anathema and time is treated as a quantum phenomenon.
However, there is no ``particle of time'' or \emph{chronon }per se. Neither
is there any operator of time. Instead, a fundamental discrete topological
measure of time occurs, called a \emph{q-tick}, or \emph{quantum tick. }The
peculiarity of the q-tick is that it has a variable conventional temporal
measure, depending on the context in which it is experienced.

\subsection{Emergence}

An \emph{emergent} quantity is something which is not in itself fundamental
but the consequence of more fundamental processes. Emergence can relate to
laws of physics, theories and conceptual structures such as continuous
spacetime. For example, the continuum theory of fluid mechanics must be an
emergent theory because fluids consist of atoms and molecules.$\;$

This leads to a principle which has been suggested in [9,10] 
and many other papers too numerous to list:

\begin{quote}
\textbf{The principle of emergence: }\emph{Because they themselves are
emergent phenomena,}\textbf{\ }\emph{humans perceive} 
(\emph{most of}) \emph{the Universe
in emergent terms.}
\end{quote}

This does not say that emergent concepts are wrong but warns about the use
of them in the formulation of fundamental laws. An example is the concept of
observer\emph{\ }in quantum mechanics. Amongst its virtues, quantum
mechanics is pre-occupied with what goes on in the laboratory and the notion
of observer is based on the actions of real physicists as they prepare
states and then perform tests on them. The problem arises because physicists
are themselves emergent phenomena. This has led to an unsatisfactory mixture
of classical and quantum concepts resulting in the measurement problem in
quantum mechanics. In this paper a more fundamental, mechanistic view of the
observer is taken.

\subsection{Consciousness}

As a corollary of the principle of emergence, consciousness has to be
recognized as an emergent phenomenon, contrary to notions currently being
taken seriously in various circles \cite{ZIZZI-01}. It cannot be an accident
that physics has made spectacular progress without incorporating
consciousness directly into any of its mechanistic laws (except for the
concept of observer in quantum mechanics). Moreover, neuroscientific
evidence exists for the idea that consciousness arises as a secondary
process following prior processing in the subconscious \cite
{HALLIGAN+OAKLEY-00}. Even if new physics were necessary to explain
consciousness, as suggested in \cite{HAMEROFF-99} and \cite{PENROSE:94},
such physics would be in accordance with the machine principle, stated
below. The laws of physics on a fundamental level must be sufficient to
account for consciousness \cite{COLLINS-01} and all other emergent
quantities.

\subsection{Quantum tests and observers}

The elimination of consciousness from fundamental physics raises the
question of the status of the observer in quantum mechanics. In the standard
formulation \cite{PERES:93}, a typical quantum experiment goes as follows.
First, at initial time $t=t_0$, a semi-classical observer prepares a system
in some initial state represented by a state vector $|\psi
\rangle $ in some Hilbert space $\mathcal{H}$. Then the system is left \emph{%
completely} alone until some later time $t=t_1>t_0,$ at which time the
observer arranges a test $\Sigma $ of the system. According to quantum
principles this will have one outcome from a number of possibilities. The
test is represented mathematically by some Hermitian operator $\hat{\Sigma}$
acting on elements of $\mathcal{H}$, and it is a fundamental postulate that
any possible outcome is an eigenstate $|\phi _k\rangle $ of this operator,
i.e., 
\begin{equation}
\hat{\Sigma}|\phi _k\rangle =s_k|\phi _k\rangle ,\;\;\;1\leq k\leq
N,  \label{456}
\end{equation}
where $s_k$ is real and represents a classical outcome of the test. It is
beyond the black arts of quantum mechanics to predict which individual
outcome will occur in any single run of a test. It is only when the
experiment is repeated many times that the relative conditional
probabilities $P\left( \phi _k|\psi \right) \equiv |\langle \phi _k|\psi
\rangle |^2$ of the various outcomes manifest themselves.

Because observers in quantum mechanics appear to have complete freedom in
deciding which states to prepare and then which tests to apply to them,
there quite naturally arises the notion that consciousness and free will
should have a role in the principles of the subject \cite{WIGNER-67}.

This may be valid on the emergent level but must be incorrect on the
fundamental level. Consider the decay of an unstable particle such as the
neutral pion. When a $\pi ^0$ is created in some particle experiment, it
will decay into one of a number of possible channels, such as $\pi
^0\rightarrow e^{+}+e^{-}$ or $\pi ^0\rightarrow \mu ^{+}+\mu ^{-}$, amongst
others. Particle tables give branching ratios for these various decays. The
decay process may be regarded as a test of the free pion state, but no
consciousness is involved in setting up this test. Its set of possible
outcomes does not appear to be determined by external factors, and certainly
not by the experimentalists. It must be determined by the fundamental laws
of the Universe. Given that there are no hidden variables, the conclusion is
that the act of preparation of a free $\pi ^0$ itself determines in some way
the test involved.

\subsection{The machine principle}

Such reflections on tests and observers lead to the following:

\begin{quote}
\textbf{The machine principle}: \emph{The Universe runs as a quantum
automaton, preparing its own states and the tests of those states. Conscious
observers are emergent complexes of states and tests and are not necessary
to the running of the Universe.}
\end{quote}

There is no need for semi-classical observers according to this principle.
The evidence is clear in the red shift of the galaxies and the fossil
record. No human observers were present in the remote past, and it is safe
to say that no other forms of consciousness doing physics experiments were
present either. Things just \emph{happened }during the normal running of the
universe.

Emergent structures such as observers are not excluded by the machine
principle, however. Consciousness is an empirical fact, as is the existence
of physicists who decide on what sort of experiments to do in their
laboratories.

The principle of emergence and the machine principle lead to a two level
view of the Universe. On the fundamental level it runs as a self-organizing
system and the spacetime continuum does not exist. On the emergent level,
the Universe forms transient patterns which appear conscious and to have
free will. These are the traditional semi-classical observers used in
discussions of orthodox quantum mechanics. These observers imagine that
spacetime is continuous, that they are embedded in it, and that they can
decide on which states to prepare and what tests to perform on them.
Fortunately, because such emergent observers emerge from quantum processes
which are inherently unpredictable, the actions of these observers are not
deterministic, though often highly predictable. Humans are not mere
automatons, for the essential reason that the running of the universe is 
\textbf{not} quite like a classical cellular automaton \cite{WOLFRAM:86}.

\subsection{Testing the future, not the past}

The eigenstates $\left\{ |\phi _k\rangle :k=1,2,\ldots N\right\} $ of a
Hermitian operator $\hat{\Sigma}$ representing a test $\Sigma $\ form a
complete set, which may be assumed orthonormal. Therefore, any state $|\psi
\rangle \;$being tested can be written as a linear superposition of those
eigenstates, 
\begin{equation}
|\psi \rangle =\sum_{k=1}^N\psi _k|\phi _k\rangle ,  \label{779}
\end{equation}
where the coefficients $\psi _k$ are complex numbers. Two points of
interpretation can be made here:

\begin{enumerate}
\item  The probability $P\left( \phi _k|\psi \right) $ of the test having
outcome $|\phi _k\rangle $ is given by 
\begin{equation}
P\left( \phi _k|\psi \right) =|\psi _k|^2.
\end{equation}
If the test is performed by an emergent observer, then the observer can
decide to perform the test many times, and home in on these probabilities in
terms of frequencies. If however the test occurs because of the machine
running of the Universe, the notion of probability is meaningful for the
very good reason that the Universe will sooner or later run through the same
test on an equivalent state a vast number of times throughout its history;

\item  A particular outcome such as $|\phi _k\rangle $ of a test can be
considered as having occurred because the initial state $|
\psi _{in}\rangle $ was in that state $|\phi _k\rangle $ all along, in a
quantum sense. The purpose of a test becomes simply to filter out this
component from the other states in the superposition $\left( \ref{779}%
\right) $.

This point of view is retrospective, in that it considers a test as an
examination of what is already there, vis., the state being tested.
\end{enumerate}

An alternative view would be to look the other way. In this view, a test
such as $\Sigma $ is something which deals with the possible future and not
the past. The role of the initial state is now simply to provide information
which helps inform $\Sigma $. This information is just one component of
perhaps a vast amount of information obtained from other events and tests
which is needed to construct or define the test $\Sigma $ of possible future
outcomes.

From this point of view, a test is more like a gateway or portal to the
future. The spectrum of eigenvalues associated with a test represents
information about that test and its possible future outcomes, and not about
the initial state being tested per se. The fact that the spectrum of
eigenvalues of an observable is independent of any state being tested is
consistent with this alternative view.

This way of looking at the process of ``measurement'' (a misnomer from this
point of view) makes sense when the question arises of incompatible tests
such as position and momentum measurement for a particle. It is not the case
that a state of a particle ``cannot have both definite position and
momentum''. It doesn't have either, as the Kochen-Specker theorem suggests 
\cite{ISHAM:95}. Rather, position and momentum tests are incompatible and so
it is not possible to have a test with an outcome which is simultaneously an
eigenstate of position and momentum. This is essentially Bohr's position on
the Einstein-Podolsky-Rosen paradox.

Any experiment in this view, therefore, is not about looking into the
properties of a state constructed in the past but about looking into the
future and providing opportunities for one of the alternatives to become
real. Nevertheless, the term \emph{test} for this process will be retained
to avoid confusion.

\subsection{Information}

Information from the active present (defined below) is used by the machine
Universe to test for the future. This information determines tests, not the
outcomes of those tests (which is the quantum part of the running of the
universe). The notion of information used here is essentially the same as
given by Deutsch \cite{DEUTSCH-01} and may be summarized as follows:

\begin{quote}
\emph{A test} $\Sigma$ \emph{contains information about an event} 
$A$ \emph{or some other test} $\Omega$ \emph{if any counterfactual
change in} $A$ \emph{or} $\Omega$ \emph{would change the possible
outcomes of} $\Sigma$ \emph{in a physically meaningful way.}
\end{quote}

Information as discussed here always comes in a classical form, which means
that it is always certain (even if unknown to some emergent observer). This
includes state vectors, which represent pure states and which are equivalent
to having a maximal amount of information in the quantum sense. Knowing that
a system is in a definite state $\Psi $ is equivalent to having a piece of
classical information.

An example of a counterfactual change in an event which would not have any
physically meaningful effect is multiplication of its event state by an
arbitrary phase. This would have no effect on the probability of any outcome
of any test of that state.

It is not the case that the only information of physical value consists of
expectation values. In the real world, a single outcome of a single
experiment gives real, physically meaningful information content. It is
possible to be sure for example that a single electron has emerged from a
Stern-Gerlach experiment in a spin up state simply by blocking off the down
beam. Such a process is known as state preparation. When more than one
outcome is possible however, which one occurs in reality is not predictable
usually from quantum mechanics. It is customary to take the view that the
only thing that matters is the observer's knowledge about the initial state,
which comes down eventually to probabilities \cite{PERES:93}. This cannot be
the entire story, as will be argued below.

\subsection{Q-ticks}

The transition from a prepared quantum state to one of its possible quantum
outcomes following a test will be defined as one tick of a fundamental
quantum clock, \emph{regardless of what the process is}. Such a tick will be
called a \emph{q-tick} (\emph{quantum tick}). For example, a neutral pion
decaying
into two photons represents one q-tick. A uranium atom decaying after ten
thousand years also represents one q-tick. A photon passing from a source
through a double slit and impinging on a screen also takes one q-tick. A
photon emitted from a quasar eleven billion years ago impinging on our
retina now takes precisely one q-tick to do so.

What determines a q-tick is irreversible information transfer, which occurs
in one of two distinct ways. First, old information from events and states
in the active present is used by the self-testing machine Universe to define
new tests of itself. Second, new information is created when quantum
outcomes of those tests occur.

If a process involves no real physical information transfer beyond a given
test to the wider Universe, then no q-tick is counted. For example, if an
outcome of some test is subsequently passed through a second, identical test
then no new information can be extracted from that second test. Therefore
this double test involves only one q-tick.

An application of this principle occurs when Feynman diagrams are used to
discuss scattering processes in elementary particle physics. Such a process
involves a single q-tick lasting from laboratory time $t=-\infty $ to time $%
t=+\infty $. The number of vertices in Feynman diagrams cannot be a
physically meaningful quantity \cite{HITCHCOCK-00}.

The essential point is not that a q-tick takes any specific externally
measurable time, but that it represents the appearance of new information
with the resolution of a single quantum outcome in some quantum test.

\subsection{The Copenhagen principle}

From the point of view of the theory being discussed here, the spacetime
continuum does not exist and quantum non-locality in time as well as space
is assumed to be meaningful. Bohr realized that there was a truly terrifying
implication of quantum mechanics: in between the preparation of a state of a
system and the testing of an outcome, the system cannot exist or be real in
any classical sense. The act of observation itself creates the reality being
observed. This leads to another fundamental principle:

\begin{quote}
\textbf{The Copenhagen principle: }\emph{Reality does not exist during a
q-tick, but only at the end of a q-tick.}
\end{quote}

The fundamental question now is, what does it mean to say that an outcome
exists? Three points of view are possible. Adherents of the many-worlds
interpretation of quantum mechanics would say that all outcomes of a test
occur, each in its own universe $\left[ 21, 22\right] $, whereas many
traditionalists would say that a state is not an objective property of an
individual system but a construct of an observer, with state reduction
taking place only in the consciousness of the observer \cite{PERES:93}.
Other traditionalists would argue that an actual outcome at the end of a
q-tick is a real physical event.

Both of the last two views are in accordance with the machine principle, and
relate to the two levels of looking at the Universe. On the emergent level,
observers deal with information and process it as they themselves evolve in
time. On the fundamental level, quantum outcomes occur physically in the
machine running of the Universe.

It does not matter that an outcome cannot be quite like a classical
measurement, because by standard quantum principles only half of available
phase space can be certain at any time. Nevertheless, \emph{something very
real occurs} in an irreversible way. A photon going through a double slit
experiment and impacting on a photographic plate does so in a very definite
part of the plate. This is what Bragg's principle means. In any discussion
of wave-particle duality, the particle aspect makes sense only \emph{after}
something has occurred, and then the wave aspect is no longer needed. Past
and future are indeed distinguished by state reduction.

Bohr's ideas are part of what is now known as the Copenhagen Interpretation.
Taken in its extreme form, this says that there is no way of influencing any
single outcome of an experiment directly, not because of any limitation on
our part, but because the outcome \emph{simply does not exist until it occurs%
}. Q-ticks are intervals of non-existence.

However, this cannot be the complete picture. Whilst there are no hidden
variables existing as a substrate of reality, information coming from the
past must somehow be causally involved in deciding the range of possible
outcomes at the end of a q-tick. This is no more radical an idea than the
use of the Schr\"{o}dinger equation to determine a future state vector, or
the Heisenberg operator equation to determine quantum operators at future
times. A differential equation is just another way of propagating classical
information available at some initial time forwards into the future, and
this information/memory occurs in the form of boundary conditions formulated
in the past. This information represents the particle aspect of Bragg's
principle. The way that it is used to predict the future concerns the wave
aspect of Bragg's principle. Moreover, when such differential equations are
discretized they appear to all intents and purposes as examples of
generalized cellular automata \cite{JAROSZKIEWICZ-97B}. None of these
deterministic models gives any statement about actually what happens at the
future end of a single q-tick, which is why the state reduction concept
appears as a blemish in an otherwise elegant picture.

Attempts to move away from the Copenhagen interpretation, such as hidden
variables theories or the many worlds interpretation, are really attempts to
avoid the conclusion that reality has no existence until resolution occurs.

\section{Discreteness}

Given that quantum processes underpin all of the properties of the Universe,
then continuous spacetime must be an emergent concept. This in turn implies
that the concept of metric and even the dimension of space are also emergent.

The notion that length represents a counting process of elementary units has
been attributed to Riemann \cite{SORKIN+al-87}. This would help solve the
problem of where a fundamental scale comes from. A counting process has no
scale.

Physicists would also like to have a dynamical explanation of why physical
space is three dimensional on emergent scales. Currently, extra dimensions
are regarded favourably by physicists, and this is consistent with the
notion discussed by Bombelli et al \cite{SORKIN+al-87} that discrete sets
with some concept of ordering (causal sets) can have emergent dimensions
which differ on different emergent scales. It is possible that the
dimensional regularization method used in the regularization of quantum
field theories works not because of some special mathematical trickery, but
because spacetime dimension really is an emergent quantity and this
particular regularization process is homing in onto this somehow.

These considerations lead to a picture of the Universe as a collection of
discrete objects called \emph{events }and discrete quantum processes called 
\emph{tests. }Events come in many varieties and so should not be visualized
as necessarily points in or of spacetime. So what are they?

Recall first a fundamental feature of quantum mechanics called \emph{%
entanglement, }which does not occur in classical mechanics.\emph{\ }In
quantum mechanics\emph{\ }it is possible to have a state which is not a
single tensor product state of more elementary vectors, such as $|\Psi
\rangle \equiv |\psi \rangle \otimes |\phi \rangle $ but an entangled one,
such as $|\Phi \rangle \equiv |\psi \rangle \otimes |\phi \rangle +|\rho
\rangle \otimes |\chi \rangle $. Such an entangled state cannot be described
as a direct product by any linear change of basis.

Given that the Universe is a quantum one, then it should be describable in
terms of some state $|\Psi \rangle $ at a given time$.$
This may be regarded as a single event. Now if the Universe were in a
completely entangled state then there would be no possibility of dividing it
in constituent parts on the fundamental level. Fortunately, the Universe
seems to be divisible into systems and observers on emergent scales and
so this property is assumed to hold at the fundamental level also.
This is the most critical assumption made in this paper. Without it no
further discussion would be possible. It is an example of Fourier's
principle of similitude \cite{JAROSZKIEWICZ-99}, applied in a quantum
context. It is conceivable, after all, that the factorization into observers
and systems on emergent scales is itself an emergent property not holding at
the fundamental level.

The state of the Universe $|\Psi \rangle $ is assumed here to factor out
into a direct product of a vast number of factor states: 
\begin{equation}
|\Psi \rangle =|\psi _1\rangle \otimes |\psi _2\rangle \otimes \ldots
\end{equation}
The factor states $|\psi _n\rangle ,\;n=1,2,\ldots $ in this product are
what is meant by \emph{events }in this paper\emph{.\ }Many of these factor
states may themselves be entangled products of even more elementary states,
but others will be fundamental themselves or even direct products. How the
event structure is factored depends on the context of the discussion, and in
a sense it does not matter.

It will be obvious from this why the discrete objects in this theory cannot
be identified with points in a discrete spacetime \emph{per se}. The matter
is more subtle than that.

Quantum processes represent the moment of the present, and so the event
structure of the Universe has a temporal aspect and may be discussed at
different stages of its temporal evolution. Emergent time splits into past,
present and future, and likewise, events fall into three categories. At a
given stage in the history of the Universe, \emph{past events} are all those
events relating to the particle aspect of Bragg's principle. These play no
further role in forming the future, at that stage. \emph{Active events} are
events which are involved in determining tests for as yet unresolved future
events and form the \emph{active present}, at the same stage. Finally, \emph{%
future events }are hypothetical events which may be outcomes of tests and do
not yet have any physical resolution, again at that stage. The temporal
status of an event therefore depends on the stage at which the Universe is
being discussed.

The discrete topological relationships between events (which were called 
\emph{links }in\emph{\ }\cite{JAROSZKIEWICZ-99})\emph{\ }are as important to
the running of the Universe as its events structure. Links represent tests
which the machine running of the Universe sets up before \emph{q-ticks }%
occur. The result of such a test is an outcome at the end of a q-tick, that
is, a resolution from a set of possible future events into a single real one.

\subsection{Diagrammatic notation}

The diagrammatic notation introduced in \cite{JAROSZKIEWICZ-99} may be used
to clarify discussions of temporal processes. The interpretation of these
diagrams is somewhat different now because quantum mechanics has been
introduced into the theory. The diagram rules are as follows:

\subparagraph{circles:}

\begin{itemize}
\item  An event $A$ in an entangled state $|\Psi \rangle \;$is represented
by a single large circle labelled internally by either $A$ or $\Psi ;$

\item  An event $B$ in a direct product state $|\psi \rangle \equiv |\phi
\rangle \otimes |\chi \rangle $ may be represented in one of four equivalent
ways: either as a single large circle labelled internally by $B$ or $\psi $,
or else as two large circles labelled $\phi $ and $\chi $ respectively or by
two convenient event labels such as $E$ and $F$ respectively;

\item  A test $\Sigma \;$is represented by a small circle labelled
internally by $\Sigma ;$

\item  A \emph{complex\thinspace \ }$O$ is a collection of events and
tests
which represents an observer, the rest of the Universe, or whatever is
factored out from a given situation and is not being tested for in the
process under consideration. A complex $O$ is represented by a large circle
labelled internally by $O$.
\end{itemize}

\subparagraph{lines:}

\begin{itemize}
\item  An event $A$ being tested by test $\Sigma $ is connected by a single
line from $A$ to $\Sigma $ with an arrow pointing from $A$ into $\Sigma ;$

\item  Two or more events being tested by test $\Sigma $ are each connected
by a single line to $\Sigma ,$ each with an arrow pointing into $\Sigma ;\;$%
equivalently, these events may be regarded as a single product event with a
single arrowed line connecting it to $\Sigma $;

\item  An outcome of a test $\Sigma $ is an event connected by a line or
lines to $\Sigma \;$with arrows pointing out of $\Sigma ;$ if the outcome is
regarded as a single event, as occurs with an entangled state, there will be
only one outcome line with an arrow. If the outcome is a direct product,
this may be represented in the same way as an entangled state above, or as
several events corresponding to the various factor states of the product. In
this case, each of these outcome products is connected by its own single
line to $\Sigma $, with arrows pointing out of $\Sigma ;$

\item  Lines without arrows represent classical information which helps
determine tests. Such lines can come from events and other tests. The
direction of information flow will be implied by the arrows in other lines
in the diagram. In any circumstance, information can flow only \emph{from }%
an event state;

\item Double lines with arrows connect complexes of events with complexes
of tests.
\end{itemize}

\subparagraph{shading:}

\begin{itemize}
\item  Events, tests and complexes which are in the active present are
shaded;

\item  Tests which are involved in informing tests with as yet unresolved
outcomes are regarded as in the active present and are therefore shown
shaded;

\item  All events, tests and complexes not in the active present are
unshaded.
\end{itemize}

\begin{figure}[t]
\begin{center}
\includegraphics[width=179.7pt,height=143.6pt]{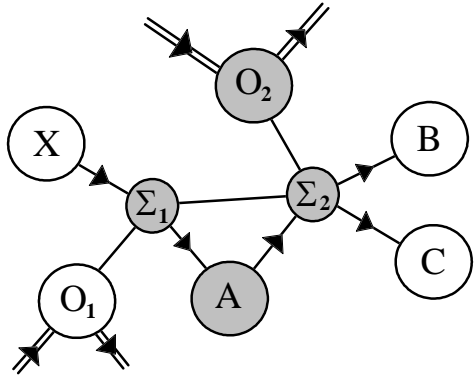}
\caption{A typical quantum process.
Arrows point from initial states to their tests and then to their outcomes,
and give the local direction of time on the q-tick level. Shaded components
belong to the active present.}
\end{center}
\end{figure}

A typical process is shown in figure $1.$ In this diagram, complex $O_1$ provides information which determines test $%
\Sigma _1$ of event $X$, with resolved outcome at event $A$. Subsequently,
complex $O_2$ and test $\Sigma _1$ determine test $\Sigma _2$ of event $A,$
with an unresolved outcome at this particular stage. It happens to be a
direct product state and hence can be represented by events $B$ and $C$.

The active present consists of tests $\Sigma _1$ and $\Sigma _2$, event $A$
and complex $O_2$, and therefore these are shaded. Event $X$ and complex
$O_1$
belong to the absolute past, whilst events $B$ and $C$ belong to the future.

Such diagrams are \emph{dynamic}, in that they depend on the stage in the
history of the Universe being described.

\subsection{Quantum automata}

The concept of a \emph{cell} is used in the theory of cellular automata
\cite
{WOLFRAM:86} as a temporally enduring container which contains a time
dependent variable. This variable is usually discrete, its value at any
given discrete time being determined by the values of the variables in
neighboring cells at earlier times. There are numerous references to
cellular automata in the literature discussing discrete spacetime.

Such cellular automata are inadequate representations of the running of the
Universe, because the events envisaged in this paper do not have any sort of
identity which propagates into the future. The Universe is more accurately
described as a quantum automaton. The main characteristic of such an
automaton is that the discrete topological relations between events and
tests in the future is uncertain. These relations depend on outcomes of
tests which have not occurred at a given stage.

\section{The Machine Observer}

Once the notion of a conscious observer has been bypassed, the questions
remain of how states are prepared and then how they are tested. It must be
the case that on a fundamental level, the Universe runs as a vast quantum
automaton. Its active present determines its own tests and then the random
outcomes of these tests become involved in a new active present, which then
determines the next set of tests. This leap-frog process proceeds ad
infinitum on a vast scale of events. Each small part of classical reality is
formed when any particular outcome is resolved (state reduction), which is
the end of one q-tick and the start of the next one, locally.

The process of time is, therefore, directly related to state reduction and
it proceeds irreversibly. It is non-integrable, in that there is no
universal clock regulating the running of the Universe. A single q-tick
could in principle last over the entire history of the Universe from the Big
Bang to the present, or it could appear to last on a Planck scale. An
analogy with the single celled organism amoeba is useful: an amoeba flows
steadily towards its food, as advanced pseudopods reach out forwards whilst
others retain their hold for a while on places where the organism had been.
Eventually the whole organism moves forwards.

Q-ticks are not involved with Schr\"{o}dinger time evolution; quite the
contrary. Schr\"{o}dinger evolution occurs in continuous time quantum
mechanics precisely in the absence of a q-tick and represents the process of
non-observation of a state of a system. From the perspective of this paper,
the time in Schr\"{o}dinger evolution is a marker of q-ticks involved with
emergent observers, not the system being observed. The possibility of
transforming to the Heisenberg picture supports this view, for in the
Heisenberg picture states do not evolve whilst operators involved in tests
do evolve.

At the end of a q-tick the following principle holds:

\begin{quote}
\textbf{The weak quantum principle: }\emph{The outcome of any test on an
event state is an eigenstate of some Hermitian operator associated with that
test}$.$
\end{quote}

This says something about the future and is a milder version of another
principle saying something about the past and the future:

\begin{quote}
\textbf{The strong quantum principle:} \emph{Every event state is an
eigenstate of some Hermitian operator representing a test.}\textbf{\ }
\end{quote}

The difference between these principles is that the strong quantum principle
rules out ``Garden of Eden'' states. These are states which could not be
explained as the outcomes of past tests. Such a concept is encountered in
the theory of cellular automata \cite{WOLFRAM:86}, and is relevant to the
origin of the Universe, which is outside the scope of the present paper.

\subsection{Null tests and unobserved phases}

In real experiments, physicists can prepare states and then decide not to
test them for arbitrary lengths of time, as measured by clocks associated
with the physicists. A typical process is shown in figure $2a$. Here a
complex of tests $\Sigma _0$ has a complex of outcomes $O_0$, which
represents an emergent observer. This constructs a test $\sigma $ which
prepares an outcome state $\psi ,$ which serves as an initial state for a
subsequent test $\Lambda $ with outcome $\phi $. The observer meanwhile runs
on to $O_1$ and then to $O_2,$ and only then constructs test $\Lambda .$

\begin{figure}[t]
\begin{center}
\includegraphics[width=317.3pt,height=229.3pt]{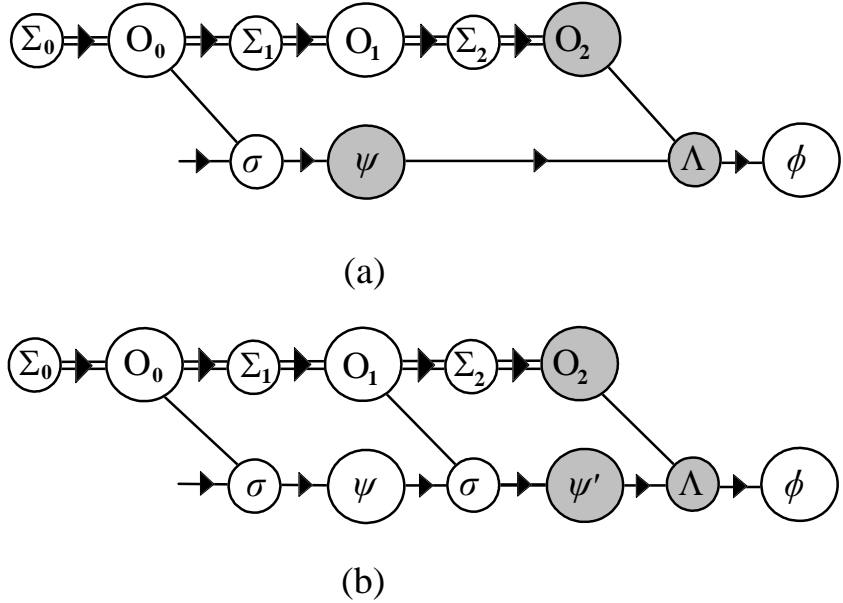}
\caption{Two equivalent
diagrams. In $2b$ a null test is included. The outcome $\psi ^{\prime }$ is
in the same ray as the initial state $\psi $ of the test.}
\end{center}
\end{figure}

In figure $2\left( b\right)$, a physically equivalent diagram suggests
that 
$O_1$ has performed a null test $\sigma $ on $\psi $, which has had no
physical effect. The outcome is $\psi ^{\prime }$ which must be proportional
to $\psi ,$ because no real information can be passed back to the observer.
A null test is physically equivalent to not doing a test on a state.

In this and other examples, the null test can be represented by the same
operator $\hat{\sigma}$ which had as outcome \thinspace $|%
\psi \rangle ,$ the initial state. The initial state is an eigenstate of $%
\hat{\sigma}$, according to quantum principles. The possible outcomes of the
null test are also eigenstates of $\hat{\sigma}$ and these differ from the
eigenstates of the original preparation test by at most random phases.
Taking into account the probabilities of the possible outcomes, which are
given by 
\begin{equation}
P\left( \psi ^{\prime }|\psi \right) \equiv |\langle \psi ^{\prime }|\psi
\rangle |^2,
\end{equation}
it is easy to see that the outcome $|\psi ^{\prime }\rangle $ of the null
test must be given 
\begin{equation}
|\psi ^{\prime }\rangle =e^{i\theta }|\psi \rangle ,  \label{556}
\end{equation}
because it occurs with probability one. All other eigenstates of the null
test occur with probability zero.

In ($\ref{556})$, the phase $\theta $ is arbitrary. No physical information
therefore can reside in this phase. This is consistent with the fact that
the initial state $|\psi \rangle $ is itself fixed only up to some arbitrary
phase.

Null tests are important because they must be involved in disentanglement in
some way. This will be discussed in the section on the EPR paradox below.

Physical null tests can always be constructed in the laboratory by real
physicists. These first prepare a state by choosing a given outcome of
some
test, which therefore serves as a filter. The equivalent of a null test is
then made by passing such a prepared state through a physically equivalent
filter. For example, an electron prepared in a spin-up state by being passed
into the spin up channel of a Stern-Gerlach apparatus with quantization axis 
$+\mathbf{k}$ will automatically pass into the spin-up channel in a second
Stern-Gerlach experiment with the same quantization axis, even if both
channels are open in the second experiment.

In this example, note that what is required for the construction of a real
null test is information about the preparation of the initial state. This
information is carried into the future by the physicists as they themselves
evolve in time, and then used to construct the second filter. In the
framework discussed in this paper, the machine Universe will have available
to it information about tests it carried out in the past, and about the
outcomes of those tests. Therefore, in principle, the machine Universe does
have all the information it requires to inform null tests on given states.
Something like this \emph{must} happen in EPR experiments.

\subsection{Example}

An example is now given of a system which constructs tests of itself which
depend on its current state and other information from the past. In this toy
model the active present always consists of a single event labelled by $E$
and indexed by an integer $n$. As the universe runs, the active event
changes from $E_n$ to $E_{n+1}$ at the end of a q-tick, and so on. The event
state $|\psi _n\rangle $ in each $E_n$ is some element of two dimensional
spin space $\mathcal{H}_2$ with basis set $\left\{ |\uparrow \rangle
,|\downarrow \rangle \right\} .$

The model starts running at time $n=0$. Now in accordance with the strong
quantum principle stated above, the event state $|\psi _0\rangle $ of the
universe at that time must be an eigenstate of some operator $\hat{\Sigma}_0$%
, which represents the net effect of whatever happened in the past in this
universe before that time, or the equivalent of the moment of the Big Bang.
This operator is taken to have the form 
\begin{equation}
\hat{\Sigma}_0\equiv \mathbf{\sigma }\cdot \mathbf{a,}
\end{equation}
where the components of $\mathbf{\sigma }$ are the Pauli matrices and $%
\mathbf{a}$ is some unit three-vector. Such an operator has only two
eigenstates, with eigenvalues $+1$ and $-1$ respectively. Hence the
following eigenvalue equation must hold: 
\begin{equation}
\hat{\Sigma}_0|\psi _0\rangle =\lambda _0|\psi _0\rangle ,
\end{equation}
where $\lambda _0=+1$ or else $\lambda _0=-1$.

It is here that classical information occurs. The eigenvalue $\lambda _0$
cannot be uncertain, even if it is unknown to any emergent observer (if such
a phenomenon were possible in this model). This information is predicated on
the nature of $\hat{\Sigma}_0$.

According to the machine principle, the state of the universe at time $t=n$
and other information from the past alone determine the test $\Sigma _{n+1}$
which will be applied to $E_n$ during the next q-tick. In the model, $\Sigma
_{n+1}$ is represented by the Hermitian operator 
\begin{equation}
\hat{\Sigma}_{n+1}\equiv \frac{_1}{^2}\left( 1+\lambda _n\right) \hat{U}\hat{%
\Sigma}_n\hat{U}^{\dagger}+\frac{_1}{^2}\left( %
1-\lambda _n\right) \hat{V}\hat{%
\Sigma}_n\hat{V}^{\dagger},\;\;\;n=1,2,\ldots
\end{equation}
where the operators $\hat{U}$ and $\hat{V}$ are elements of $SU\left(
2\right) $. These operators are regarded here as generated by the laws of
physics in this particular universe with no further explanation as to their
origin. The diagram for the running of this toy universe is given in figure. 
$3$:

\begin{figure}[t]
\begin{center}
\includegraphics[width=256pt,height=104pt]{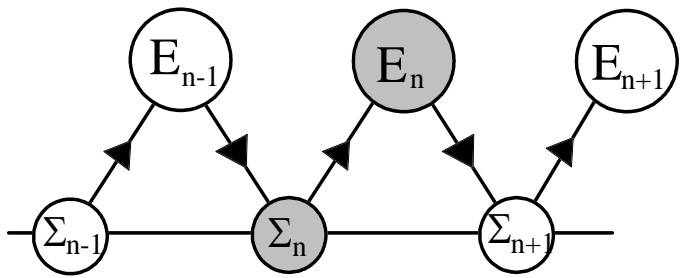}
\caption{The running of the given toy universe.}
\end{center}
\end{figure}

The test $\Sigma _{n+1}$ represented by operator $\hat{\Sigma}_{n+1}$
requires information from the immediate past, not only about the state of
the event $E_n$ but also about the test $\Sigma _n$ which led to it. Without
all of this information the eigenvalue $\lambda _n$ is undetermined and so $%
\hat{\Sigma}_{n+1}$ cannot be constructed by the machine.

Assuming that information about $\lambda _n$ is available, then $\hat{\Sigma}%
_{n+1}$ can be constructed by the machine universe. The eigenvalues of this
operator are always $\pm 1$, so that the future is uncertain before the next
q-tick. At the end of that q-tick, however, the active present of the
Universe shifts to event $E_{n+1}$, with $|\psi _n\rangle $ collapsing onto
one of the two possible outcomes of the eigenvalue equation 
\begin{equation}
\hat{\Sigma}_{n+1}|\psi _{n+1}\rangle =\lambda _{n+1}|\psi _{n+1}\rangle ,
\end{equation}
where $\lambda _{n+1}=\pm 1.\;$

In the model, the operators $\hat{U}$ and $\hat{V}$ may be assumed
incommensurate, i.e., there exist no positive integers $p,q$ such that $\hat{%
U}^p=\hat{V}^q.\;$Then given $|\psi _0\rangle ,$ $\lambda _0$ and $\hat{%
\Sigma}_0$ there are four distinct possible states of this universe at the
end of the second q-tick, $n=2$. These are characterized by the four
possible histories given in terms of the eigenvalues $\left( \lambda
_1,\lambda _2\right) $, vis; 
\begin{equation}
\left( \lambda _1,\lambda _2\right) =\left( +1,+1\right), \;%
\left( +1,-1\right), \;	 \left( -1,+1\right), \; %
\left( -1,-1\right) .
\end{equation}
Which one of these possibilities is taken is fundamentally a quantum process
and cannot be predicted or forced. Starting from one state at time zero, the
number $N_n$ of distinct possible branches of the universe is $2^n$ at time $%
n$, but only one of them will occur then.

In this model, the past is unique whereas the future is uncertain. There is
only one state at any given time, so that the present is unique also.
Because outcomes are quantum processes, however, the past cannot be uniquely
retrodicted from information about the present.

In this model, time runs physically because classical information about the
eigenvalue $\lambda _n$ has to be extracted at the end of each q-tick in
order to determine the next test. An outcome has to occur for time to run,
in other words. 

\subsection{The Einstein-Podolsky-Rosen paradox}

The principles applied to elementary quantum processes can also be applied
when emergent observers are involved. These are complexes of events and
tests which to all intents and purposes can be treated diagrammatically as
single events. Because the emphasis here is on tests as much as on the
states being tested, the approach provides some insight into how
conventional measurements are made on emergent scales.

The Einstein-Podolsky-Rosen thought experiment \cite{EPR} deals with
entangled states. When certain observations are made on such states,
non-classical consequences can follow which strongly support the view that
quantum processes are non-local. This has been reinforced by experiments on
Bell inequalities and supports the approach taken in this paper, which does
not assume space exists at a fundamental level.

The version of the EPR scenario discussed here is the spin half bound state
example favoured by Bohm. Consider the creation of a neutral pion $\pi ^0$\
at time $t=0$ in some inertial frame $\mathcal{F}$ and its subsequent decay
into an electron-positron pair. The total spin $s$ of the state remains zero
during the decay, so its spin structure may be considered to be that of two
spin half particles in the entangled form 
\begin{equation}
|\pi \rangle \equiv \frac{_1}{^{\sqrt{2}}}\left\{ |+\mathbf{k}\rangle
_e\otimes |-\mathbf{k}\rangle _p-|-\mathbf{k}\rangle _e\otimes |+\mathbf{k}%
\rangle _p\right\}  \label{389}
\end{equation}
relative to the standard tensor product space basis, where $\mathbf{k}$ is a
unit vector along the $z$-direction quantization axis and the subscripts $e$
and $p$ refer to the electron and positron respectively \cite{PERES:93}.

If an observer $O$ subsequently decided to test the spin of the system, such
a test would be represented by the operator 
\begin{equation}
\hat{\Sigma}\left( \mathbf{a}\right) \equiv \left( \mathbf{\sigma }_e\mathbf{%
\cdot a}\right) \otimes \hat{I}_p+\hat{I}_e\otimes \left( \mathbf{\sigma }_p%
\mathbf{\cdot a}\right) ,  \label{888}
\end{equation}
where $\mathbf{\sigma }\equiv \left( \sigma ^1,\sigma ^2,\sigma ^3\right) $
are the Pauli matrices, $\mathbf{a}$ is a unit vector pointing in some
direction chosen by $O$ \emph{after }the state has been prepared\emph{\ }and 
$\hat{I}_e$ and $\hat{I}_p$ are identity operators in their respective
component spaces of the tensor product space.\emph{\ }Factors of $\frac{_1}{%
^2}\hbar   $ will be ignored here. This process is represented by figure $4$.

In this diagram, the initial state-event of the pion is represented by the
circle labelled $\pi $. The line with an arrow from the left of this event
comes from some test of which the pion was a particular outcome and which is
not shown. The circle labelled $O$ represents an emergent observer involved
with the test of the pion labelled\thinspace $\Sigma \left( \mathbf{a}%
\right) $.\ The double line with an arrow going into $O$ implies that $O$ is
itself the result of a large, possibly vast number of outcomes of tests in
the immediate past, and these are not shown. The observer $O$ is a complex
of elementary events, that is, $O$ is an emergent process, and for
convenience has been shown as one circle. It is a feature of the present
formulation that no part of $O$ is an unresolved outcome of a test; state
resolution (reduction) has definitely occurred in each component event
making up $O$. The observer may also consist of various tests, which are not
quantum objects themselves. Therefore overall, the observer is
semi-classical.

\begin{figure}[t]
\begin{center}
\includegraphics[width=134.1pt,height=113.8pt]{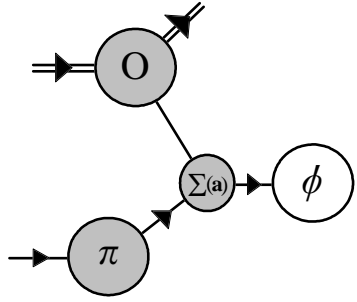}
\caption{A single local
test $\Sigma \left( \mathbf{a}\right) $ on a neutral pion state $\pi $, with
outcome $\phi .$ O represents a complex of events making up the emergent
observer.}
\end{center}
\end{figure}

The line without an arrow connecting $O$ with the test $\Sigma \left( 
\mathbf{a}\right) $ indicates that a vast amount of information from $O$ may
be involved in setting up this test, whereas a single line with an arrow
from $\pi $ to $\Sigma \left( \mathbf{a}\right) $ represents a test of that
state. The outcome of the test $\Sigma \left( \mathbf{a}\right) $ is shown
as a future event $\phi $, that is, one not yet resolved, and is therefore
an unshaded circle according to the notation.

It is a particular property of the state $\left( \ref{389}\right) $ that it
may be written in the alternative form 
\begin{equation}
|\pi \rangle =\frac{_1}{^{\sqrt{2}}}\left\{ |+\mathbf{a}\rangle _e\otimes |-%
\mathbf{a\rangle }_p-|-\mathbf{a}\rangle _e\otimes |+\mathbf{a}\rangle
_p\right\} ,  \label{299}
\end{equation}
where the component states such as $|+\mathbf{a}\rangle _e$ are eigenstates
of the component operators $\mathbf{\hat{\sigma}}_e\mathbf{\cdot a}$ in $%
\hat{\Sigma}\left( \mathbf{a}\right) $ respectively. It is readily seen that
state $\left( \ref{299}\right) $ is an eigenstate of $\hat{\Sigma}\left( 
\mathbf{a}\right) $ with eigenvalue zero, confirming that the original state
is spinless.

The $EPR$ paradox arises when two different and widely separated emergent
observers $O_1$ and $O_2$ each decide to perform their own experiment on
just one of the constituent particle spins. This is possible here because $%
O_1$ could filter out the positron because of its positive electric charge
and test only for the electron spin (say), and similarly $O_2$ could filter
out the electron and test only for positron spin. It is at this point that
free will appears to enter into the picture, which is the source of the
problem.

Observers $O_1$ and $O_2$ will consist of enormous patterns of events and
tests which on emergent scales have consciousness and the belief structures
that they are at rest and widely separated in the same inertial frame with
their conventional clocks synchronized. Suppose this is the case, and now
suppose further that on one side of the Universe, observer $O_1$ decides at
emergent (co-ordinate) time $t=T>0$ to perform a Stern-Gerlach experiment on
the electron only whereas on the other side of the Universe observer $O_2$
decides at the ``same time'' $t=T$ to perform a Stern-Gerlach experiment on
the positron only.

Observer $O_1$ therefore performs a test $\Sigma _1$ which they believe is
described by the operator 
\begin{equation}
\hat{\Sigma}_1\equiv \mathbf{\hat{\sigma}}_e\cdot \mathbf{b,}  \label{896}
\end{equation}
acting on the electron state space only, whereas observer $O_2$ performs a
test $\Sigma _2$ which they believe is described by the operator $\hat{\Sigma%
}_2\equiv \mathbf{\hat{\sigma}}_p\mathbf{\cdot c}$ acting on the positron
state space only.\ Here $\mathbf{b}$ and $\mathbf{c}$ are unit vectors
chosen by observers $O_1$ and $O_2$ respectively, with apparently full
freedom to choose any directions for these vectors\emph{.}

This freedom is the source of the problem. Suppose $O_1$ chooses axis $%
\mathbf{b}$.\ There are only two possible outcomes of the test $\Sigma _1.\;$%
\textbf{Either} the electron has spin $|+\mathbf{b}\rangle _e$ \emph{and
therefore the positron} \emph{must be in state} $|-\mathbf{b\rangle }_p,$ 
\textbf{or else} the electron has spin $|-\mathbf{b}\rangle _e$ \emph{and
therefore the positron must be in state} $|+\mathbf{b\rangle }_p$. This is
because total angular momentum is conserved.

This means that regardless of where in the Universe observer $O_2$ is, their
choice of test $\Sigma _2$ will have outcomes apparently dictated by the
outcome of test $\Sigma _1.\;$This leads to certain predictions involving
Bell inequalities and ultimately to a conflict with Einstein locality \cite
{PERES:93}, and is the source of endless debate in quantum mechanics.

\begin{figure}[t]
\begin{center}
\includegraphics[width=136.0pt,height=184.0pt]{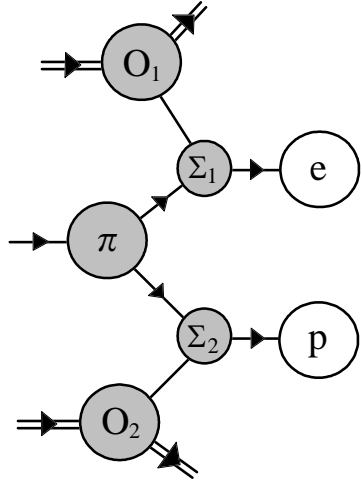}
\caption{A diagram such as this is
forbidden because $\pi $ is entangled.}
\end{center}
\end{figure}

In a diagrammatic representation of these observations, the classical and
erroneous \emph{EPR} picture of what is going on would represented by figure 
$5$. This suggests that separate parts of the entangled system are
being tested
separately, and this is the source of the \emph{EPR} problem.
According to quantum principles, a null test 
yields no new information and therefore leaves a state essentially
unchanged, whereas the extraction of new information
about a state must alter it. Therefore, only one non-null test of a state is
possible. The only circumstance where the equivalent of two independent
tests on the same event state is possible is if that state were a direct
product, and under those circumstances, one test would have to involve one
factor of the product states and the other test would have to involve
another factor. This is not possible when the initial state is entangled, as
in the present case. There is therefore a veto on diagrams such as figure $5$
when $\pi $ is entangled.

This leads to the following principle:

\begin{quote}
\textbf{The entanglement principle:} \emph{An entangled event state can be
tested by only one test, whereas each factor state of a direct product event
state can be tested separately.}
\end{quote}

This principle permits a discussion of the Universe in terms of observers
and systems, because the state of the Universe appears not to be completely
entangled.

The entangled nature of the initial pion state requires an alternative
diagrammatic description of the above experiment. Given that continuous time
and space do not exist per se and that quantum processes occur over single
q-ticks, the running of the Universe must take one of two topologically
distinct patterns of tests, shown in figures $6a$ and $6b$ respectively:

\begin{figure}[t]
\begin{center}
\includegraphics[width=303.9pt,height=208.2pt]{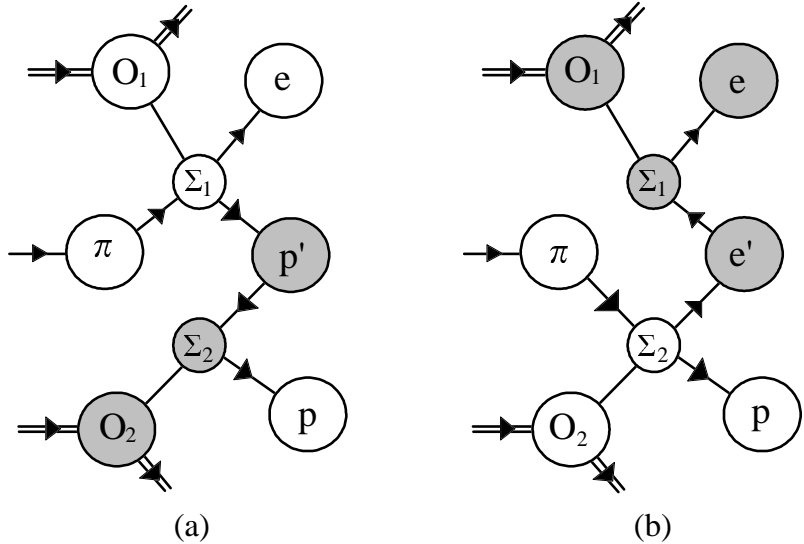}
\caption{The two
alternative and exclusive topologically distinct combinations of tests which
measure individual spins in an entangled state at widely separated places in
the Universe. A combined test takes two q-ticks.}
\end{center}
\end{figure}

The interpretation of figure $6a$ is the following. A $\pi ^0$ state is
created at the event labelled $\pi $. This state is not only an eigenstate
of the operator $\left( \ref{888}\right) $ but also of the angular momentum
multiplet operator 
\begin{equation}
\hat{S}^2\equiv \sum_{i=1}^3\left( \hat{\sigma}_e^i\otimes \hat{I}_p+\hat{I}%
_e\otimes \hat{\sigma}_p^i\right) \left( \hat{\sigma}_e^i\otimes \hat{I}_p+%
\hat{I}_e\otimes \hat{\sigma}_p^i\right) ,  \label{999}
\end{equation}
with eigenvalue zero. Although the creation of the pion is the outcome of a
single test not shown in figure $6$, the operator $\left( \ref{888}\right) $
on its own is insufficient to pin down the state of the pion and represent
that test fully.

When observer labelled $O_1$ decides to test for the electron spin with test
given by $\left( \ref{896}\right) $, $O_1$ is apparently \emph{not }testing
anything about the positron. So it could be reasonably be asked, what forces
the positron into an opposite spin state to the electron? The answer is
given by the machine principle, which says that the automatic running of the
machine universe itself ensures the positron comes out from $\Sigma _1$ in
such a way as to ensure total spin conservation. In other words, some
additional component to the test $\Sigma _1$ must be involved.

The real test $\Sigma _1$ therefore must require any outcome to be a
simultaneous eigenstate of the two operators 
\begin{eqnarray}
\hat{\Sigma}_1^{(1)} &=&\left( \mathbf{\hat{\sigma}}_e\cdot \mathbf{b}%
\right) \otimes \hat{I}_p \\
\hat{\Sigma}_1^{\left( 2\right) } &=&\left( \mathbf{\hat{\sigma}}_e\mathbf{%
\cdot b}\right) \otimes \hat{I}_p+\hat{I}_e\otimes \left( \mathbf{\hat{\sigma%
}}_p\mathbf{\cdot b}\right) .
\end{eqnarray}
A convenient basis $\mathcal{B}$ for solutions to this problem is given by
the direct products 
\begin{equation}
\mathcal{B}=\left\{ |+\mathbf{b\rangle }_e\mathbf{\otimes |+b\rangle }_p%
\mathbf{,\,|+b\rangle }_e\mathbf{\otimes |-b\rangle }_p\mathbf{,\,|-b\rangle 
}_e\mathbf{\otimes |+b\rangle }_p\mathbf{,\,|-b\rangle }_e\mathbf{\otimes
-b\rangle }_p\right\} ,
\end{equation}
where 
\begin{equation}
\mathbf{\hat{\sigma}}_e\cdot \mathbf{b}|+\mathbf{b\rangle }_e=+|+\mathbf{%
b\rangle }_e,
\end{equation}
and so on. Eigenstates of the first operator $\hat{\Sigma}_1^{(1)}$ are of
the form 
\begin{eqnarray}
|+\rangle &=&\alpha |+\mathbf{b\rangle }_e\mathbf{\otimes |+b\rangle }%
_p+\beta \mathbf{|+b\rangle }_e\mathbf{\otimes |-b\rangle }_p,  \nonumber \\
|-\rangle &=&\gamma \mathbf{\,|-b\rangle }_e\mathbf{\otimes |+b\rangle }%
_p+\delta \mathbf{\,|-b\rangle }_e\mathbf{\otimes |-b\rangle }_p,
\label{356}
\end{eqnarray}
with eigenvalues $\lambda ^{\left( +\right) }=+1$ and $\lambda ^{\left(
-\right) }=-1$ respectively. The information about which eigenvalue occurs
is transmitted to the emergent observer $O_1$ and is interpreted as an
electron in one of the two possible spin states $|+\mathbf{b\rangle }_e$ or $%
|-\mathbf{b\rangle }_e\mathbf{.}$

The coefficients $\alpha ,\,\beta ,\,\gamma $ and $\delta $ in $\left( \ref
{356}\right) $ are not arbitrary and are determined by the requirement that
the states $|+\rangle $ and $|-\rangle $ are also eigenstates of $\hat{\Sigma%
}_1^{\left( 2\right) }$ \textbf{and} the crucial requirement that no new
information is extracted from that sub-test by either $O_1$ or the machine
Universe. This means that $|+\rangle $ and $|-\rangle $ must each have
eigenvalue $zero$ relative to $\hat{\Sigma}_1^{\left( 2\right) }.$ In
effect, $\hat{\Sigma}_1^{\left( 2\right) }$ represents the action of a null
test on the components of the entangled state. This requirement fixes the
coefficients in $\left( \ref{356}\right) $ giving 
\begin{eqnarray}
|+\rangle &=&\mathbf{|+b\rangle }_e\mathbf{\otimes |-b\rangle }_p,  \nonumber
\\
|-\rangle &=&\mathbf{\,|-b\rangle }_e\mathbf{\otimes |+b\rangle }_p,
\label{478}
\end{eqnarray}
ignoring arbitrary overall phases. These occur with probability one half
each. Note that the test $\Sigma _1$ has an outcome which is not an
eigenstate of the multiplet operator $\left( \ref{999}\right) .$

Because observer $O_1$ has acquired information from test $\Sigma _1$, a
single q-tick has definitely occurred then. However, $O_1$ will almost
certainly believe that a much greater time has elapsed than just one q-tick,
because according to them, their local time is measured by their own
internal processes, which take place over vast numbers of q-ticks. These are
not shown but implied by the double lines entering and leaving the circle
representing $O_1.$

Because the outcome of test $\Sigma _1$ is a product state, it can be shown
as two events, labelled $e$ and $p^{\prime }$ in figure $6a$. The event
state $|p^{\prime }\rangle \;$is one of the positron states $\mathbf{%
|+b\rangle }_p,\mathbf{|-b\rangle }_p$ in $\left( \ref{478}\right) \,$and
this now feeds into test $\Sigma _2,$ which has been set up by observer $O_2$
and who believes themselves to be on the other side of the universe to $O_1$%
. This test will have as outcome event $p$, which occurs on the second
q-tick after the creation of the initial state $\pi $.

Even though $O_1$ and $O_2$ may believe that they are separated by vast
distances, the quantum processes given in figure $6a$ do not have any
cognisance of this. These distance estimates are particular emergent
attributes of the Universe calculated by the emergent observers via
relatively straightforward counting processes, in line with Riemann's idea
that distance is a numerical count of fundamental units \cite{SORKIN+al-87}.

The entire process could follow an alternative path, given by figure $6b.$
The two diagrams, figures $6a$ and $6b$ suggest that such experiments cannot
be carried out absolutely simultaneously in terms of q-ticks at different
parts of the universe. This is nothing to do with relativity at all. One of
the tests must always be one q-tick later than the other. This is a
topological relationship between outcomes of tests, and in that sense
involves the structure of spacetime. Co-ordinate or laboratory times
estimated by the two emergent observers relate to emergent descriptions of
the Universe. They may appear to be simultaneous to all intents and
purposes, because they count time in terms of vast numbers of q-ticks, most
of which may be thought of as occurring on Planck scales.

It is conceivable that one day technology might be found to establish
whether these topological structures are relevant in such processes, but it
likely that there might be no way in principle of determining whether
process $6a$ or $6b$ had occurred in an actual run of the experiment.

It is possible that the choice of which of these processes is actually taken
by the running of the universe may itself be thought as the outcome of some
higher order quantum test. This would perhaps be equivalent to ``second
quantization'', i.e., a quantum test whose outcomes are themselves different
tests, rather than states. Since these tests involve different topologies,
as in figures $6a$ and $6b$, there is here a scenario for an approach to
quantum spacetime topology, otherwise known as quantum gravity. That is
outside the scope of this paper and is a matter reserved for the future.

Discussions of Bell inequalities will remain unchanged in the approach
discussed here, because all the standard correlations of quantum mechanics
will be reproduced. What the diagrammatic approach taken here has done is to
emphasize why these correlations should occur. These inequalities deal with
expectation values, so they refer to emergent processes. These can be dealt
with straightforwardly here because the machine principle does not excluded
the emergent level. The use of emergent observers in the discussion
validates a discussion of probabilities, because these observers can make
the choice of repeating experiments such as the one discussed above.

\section{Particle decays and scattering}

These ideas can be applied to particle decays and scattering processes.
Consider the former. The question of particle decay lifetime involves a
balance of heuristics and formal theory, because the interpretation of what
is going on lies on the borderline between the classical and the quantum.
According to the notion of a q-tick, an elementary particle decay involves
one q-tick, whereas conventional time is measured on emergent scales and
normally involves vast numbers of q-ticks. The question arises as to what
the
physical meaning of decay lifetime is.

The answer comes from the Heisenberg picture in quantum mechanics. In this
picture, a state remains frozen in time and all the time dependence is
transferred onto the observables, the physical operators of the theory. This
is in line with the approach taken in this paper. Observables represent
tests, and their time dependence is a manifestation of time as it runs in
the Universe external to the state being observed.

It is useful to review briefly the usual approach to particle decays from
both the Schr\"{o}dinger and Heisenberg pictures.

\subsection{The Schr\"{o}dinger picture}

In this picture an initial state $|\psi \rangle $ is prepared at time $%
t_i\;$and allowed to evolve quantum mechanically until a final time $%
t_f>t_i$, at which time the evolved state \thinspace $|\psi ,t_f\rangle$%
\thinspace is tested to see if it has decayed. Suppose the decayed state
looked for is $|\phi \rangle$. This state may be assumed an eigenstate
of some operator $\hat{\Lambda}$, vis, 
\begin{equation}
\hat{\Lambda}|\phi \rangle =\lambda |\phi \rangle .  \label{666}
\end{equation}
Then the probability $P\left( \phi |\psi \right) $ that the outcome of
the test is $|\phi \rangle $ is just 
\begin{equation}
P\left( \phi |\psi \right) =|\langle \phi |\hat{U}\left( t_f,t_i\right)
|\psi \rangle |^2,  \label{199}
\end{equation}
where $\hat{U}\left( t_f,\,t_i\right) $ is the time evolution operator. In
this picture, the initial state is assumed to change in time according to
the rule 
\begin{equation}
|\psi \rangle \rightarrow |\psi ,t_f\rangle \equiv \hat{U}\left(
t_f,t_i\right) |\psi \rangle .
\end{equation}
The probability $P\left( \phi |\psi \right) $ is then transformed via a
conventional heuristic formalism into the decay lifetime associated with the
transition $|\psi \rangle \rightarrow |\phi \rangle $

A particularly subtle point is this. In this scenario, there is a vast
amount of information assumed about the time evolution and the measurement
process, which is understood by experimentalists and theorists intuitively,
but which is not mathematically incorporated into the formalism \emph{per se.%
} The physical interpretation of the probability $P\left( \phi |\psi
\right) $ is that it is the probability that a test made at time $t_f$
for the occurrence of state $|\phi \rangle $ had a positive outcome, \textbf{%
given that no attempt had been made to look at the initial state up to that
time and extract information about it}. This is classical information
about
what the \textbf{observers} have done or not done during the time interval $%
\left[ t_i,t_f\right]$.

In other words, even in the Schr\"{o}dinger picture, which gives the
impression that the only time evolution occurring lies with the state, the
behaviour of the observer in time is just as crucial. Knowing that a test
has \textbf{not }yet been done on a system is a piece of information which
is just as important in a quantum measurement as knowing that a test has
been done.

\subsection{The Heisenberg picture}

This picture is fully consistent with the notion of a q-tick. In this
picture, all time dependence is transferred explicitly onto observers and
tests. Once prepared, an initial state is frozen until it is tested. This is
much more natural a picture in terms of quantum principles than the
Schr\"{o}dinger picture. It could be argued that latter is somewhat
inconsistent, because a prepared state which is not being tested is
effectively decoupled from the Universe, so how could any sort of notion of
time be associated with it? How does an isolated state have any information
about co-ordinate time, which is measured by on observer decoupled from that
state?

In the Heisenberg picture,\ time runs for observers, and when they construct
tests this time dependence is encoded in those tests. According to this
picture, the decay experiment discussed above is described as follows. First
the observer constructs the state $|\psi \rangle $ at time $t_i$. This
state is left completely alone until time $t_f>t_i,$ at which time the
observer tests for the probability that the initial state has component $%
|\phi ,t_f\rangle .$ This state is an eigenstate of the operator $\hat{%
\Lambda}\left( t_f\right) \equiv \hat{U}^{\dagger}\left( %
t_f,\,t_i\right) %
\hat{\Lambda}%
\hat{U}\left( t_f,t_i\right) $ with the same eigenvalue $\lambda \;$as
in $\left( \ref{666}\right) $. It is the eigenvalue which identifies to
the
observer the state being measured. A solution is given by 
\begin{equation}
|\phi ,t_f\rangle =\hat{U}^{\dagger}\left( t_f,t_i\right) |\phi \rangle ,
\end{equation}
and so the probability of the transition occurring is exactly the same as
for the Schr\"{o}dinger picture, equation $\left( \ref{199}\right) .$

\subsection{The q-tick picture}

From the point of view of the picture given in this paper, a decay
experiment can be represented by a diagram such as figure $7.$

\begin{figure}[t]
\begin{center}
\includegraphics[width=400pt,height=90.0pt]{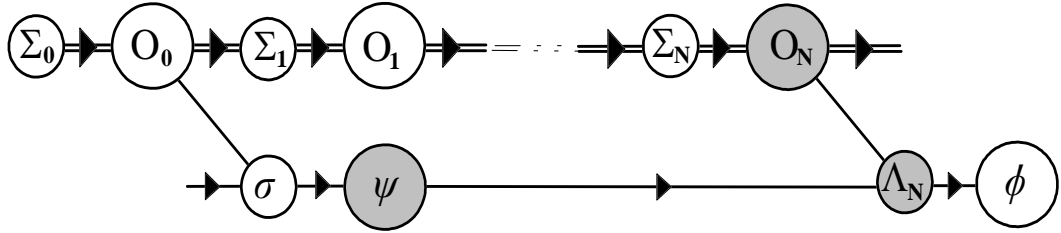}
\caption{The q-tick picture
in a decay or scattering process as seen on the fundamental and emergent
levels.}
\end{center}
\end{figure}

Here the observer runs as a quantum automaton from its initial configuration 
$O_0$ to some final configuration $O_N$, via a vast number of q-tick
processes, characterized by some integer $N$, whereas the decays process
takes precisely one q-tick on the fundamental level. Almost certainly, $N$
would not be precise, not because of any uncertainty in the topology of
events, but because time on a fundamental level in not integrable. However,
on emergent scales, estimates of the average number of q-tick processes
should be meaningful and it is these estimates which translate into
estimates by emergent observers into their laboratory time. There is no
fundamental scale per se in this framework, but if typical q-ticks
associated with emergent observers were counted to be more-or-less
equivalent to Planck units, then the decay of a $\pi ^0$ takes about $%
10^{27} $ of these observer q-ticks, i.e., $N\sim 10^{27}.$

The test performed by $O_N$ will be represented by some operator $\hat{%
\Lambda}_N$, which carries information supplied by $O_N$. If $O_N$ forms
part of a sequence of observer events $O_1,O_2,\ldots $ which is homogeneous
in the sense that no radical changes occur in the laboratory environment
associated with each element of this sequence, than the simplest ansatz
which could be applied would be to take 
\begin{equation}
\hat{\Lambda}_N=\hat{U}^{\dagger N}\hat{\Lambda}U^{N},
\end{equation}
where $\hat{U}$ is an elementary unitary time-step operator of the form
discussed in \cite{JAROSZKIEWICZ-97A}. The picture which emerges corresponds
precisely to the Heisenberg picture in the continuous time formulation.

\section{Discussion}

A number of fundamental issues remain to be discussed in this framework, and
these will be considered elsewhere. They include irreversibility, the early
machine Universe, Schr\"{o}dinger's cat, the quantum Zeno effect,
superselection rules and unbounded operators such as position and momentum.
It is likely that the emphasis placed here on tests as much as on their
outcomes should have a lot to say about some of these issues. For example,
if the strong quantum principle holds, then the issue about the origin of
the Universe really concerns the test which produced the initial state, and
not so much that initial state. Likewise, the Schr\"{o}dinger's cat issue
really concerns the physical existence or not of a test which could produce
a state which was a linear combination of a living cat and a dead cat. If
such a test cannot be constructed physically, then the question of such a
linear superposition does not arise. The same remark holds for
superselection rules forbidding linear combinations of states with different
electric charges.

On the physical meaning of time, the position taken in
this paper is that it is a real phenomenon and that the Universe\emph{\
runs}. This contrasts with some
emergent theories such as general relativity, where general covariance
leads to a
view which is unsettling and counter intuitive. Time appears to freeze out
in such theories when quantisation is attempted. 
It should
be noted that such a result can be found in any
Lagrangian model using Dirac's reparametrisation method \cite{DIRAC:64}
involving constraint mechanics, but that does not eliminate the physical
reality of time in such models.

Since time as discussed in this paper is not directly related to an
integrable parameter except on emergent scales, an obvious conclusion is
that the
Euclidean formulation of time, wherein real time is rotated into the
imaginary axis, is meaningful only in an emergent
context. 
Theories which use it exclusively cannot be truly fundamental. Lattice
gauge
theories formulated on four dimensional Euclidean lattices are reasonable
because this approach is regarded as an approximation
method. Cosmological
theories which are formulated exclusively in Euclidean spacetimes must
be regarded as unphysical from the point of view of this paper, as are 
attempts
to regularize quantum gravity by appealing to imaginary time. 

Finally, because in the framework discussed here time is the
acquisition of
quantum information, there is no scope here for closed timelike curves,
and any
theory which permits them must be emergent and therefore not fundamental.

\subparagraph{Acknowledgments:}

I am indebted to Drs Rosolino Buccheri and Vito Di Ges\`{u} and the other
organizers and participants of the \emph{First International
Interdisciplinary Workshop: Studies on the Structure of Time: from Physics
to Psycho}(\emph{Patho})\emph{logy}, held in Palermo in November $1999$,
for giving me the
stimulation and motivation to carry out this work. I am grateful to Keith
Norton and Jon Eakins for all their views, comments and discussions with me
on the nature of time.

\newpage\


\begin{thebibliography}{99}
\bibitem{SORKIN+al-87}  D. Meyer, L. Bombelli, J. Lee and R. Sorkin,
\emph{%
Space-Time as a Causal Set}, Phys. Rev. Lett., 59(5): 521--524, 1987.

\bibitem{MARKOPOULOU-00}  Fotini Markopoulou, \emph{Quantum causal histories}%
, Class. Quant. Grav., 17: 2059--2072, 2000.

\bibitem{HITCHCOCK-99}  Scott Hitchcock, \emph{Quantum clocks and the origin
of time in complex systems}, gr-qc/9902046, pages 1--18, 1999.

\bibitem{REQUARDT-00}  Manfred Requardt, (\emph{Quantum}) \emph{spacetime
as a
statistical geometry of lumps in random networks}, Class. Quant. Grav,
17:2029--2057, 2000.

\bibitem{ZIZZI-01}  P.~A. Zizzi, \emph{The early universe as a quantum
growing network}, gr-qc/0103002, pages 1--16, 2001.

\bibitem{BRAGG'S-PRINCIPLE}  Lawrence Bragg, quote attributed to Bragg in 
\emph{''Dynamical Solution to the Quantum Measurement Problem, Causality,
and the Paradoxes of the Quantum Century''}, by V.P. Belavkin, Open Sys. and
Information Dyn. 7: 101-129, 2000.

\bibitem{WHITROW:80}  G~J Whitrow, \emph{The Natural Philosophy of Time},
Clarendon Press, Oxford, 2nd Edition, 1980.

\bibitem{OMNES:94}  Roland Omn\`{e}s, \emph{The Interpretation of Quantum
Mechanics}, Princeton University Press, Princeton, New Jersey, 1994.

\bibitem{SORKIN-99}  D.P. Ridout and R.D. Sorkin, \emph{A Classical
Sequential Growth Dynamics for Causal Sets}, gr-qc/9904062, pages
1--28, 1999.

\bibitem{REQUARDT-99}  Manfred Requardt, \emph{Space-time as an
orderparameter manifold in random networks and the emergence of physical
points}, gr-qc/99023031, pages 1--40, 1999.

\bibitem{HALLIGAN+OAKLEY-00}  Peter Halligan and David Oakley, \emph{%
Greatest myth of all}, New Scientist, 18 November, pages 34--39, 2000.

\bibitem{HAMEROFF-99}  S. Hagan, S. R. Hameroff, J. A. Tuszy\`{n}ski,
\emph{%
Quantum Computation in Brain Microtubules? Decoherence and Biological
Feasibility, }quant-ph/0005025, pages 1-10, 2000

\bibitem{PENROSE:94}  Roger Penrose, \emph{Shadows of the Mind}, Oxford
University press, 1994.

\bibitem{COLLINS-01}  John~C. Collins, \emph{On the compatibility between
Physics and intelligent organisms}, http://xxx.lanl.gov/physics/0102024,
2001.

\bibitem{PERES:93}  Asher Peres, \emph{Quantum Theory: Concepts and Methods}%
, Kluwer Academic Publishers, 1993.

\bibitem{WIGNER-67}  E~P Wigner, \emph{Remarks on the mind-body question},
Symmetries and Reflections, pages 171--184, 1967, reprinted in \emph{Quantum
Theory and Measurement}, edited by J A Wheeler and W H Zurek,Princeton
University Press, 1983.

\bibitem{WOLFRAM:86}  Stephen Wolfram, \emph{Theory and Applications of
Cellular Automata}, World Scientific, 1986.

\bibitem{ISHAM:95}  C. J. Isham, \emph{Lectures on Quantum Theory}, Imperial
College Press, 1995.

\bibitem{DEUTSCH-01}  D.~Deutsch, \emph{The structure of the multiverse}, 
quant-ph/0104033, pages 1--21, 2001.

\bibitem{HITCHCOCK-00}  Scott Hitchcock, \emph{Feynman clocks, causal
networks, and the origin of hierarchical 'arrows of time' in complex
systems. part i. 'conjectures'}, gr-qc/0005074, pages 1--50, 2000.

\bibitem{DeWITT+GRAHAM:73}  Bryce~S. DeWitt and Neill Graham, \emph{The
Many-Worlds Interpretation of Quantum Mechanics}, Princeton University
Press, 1973.

\bibitem{DEUTSCH:97}  David Deutsch, \emph{The Fabric of Reality}, The
Penguin Press, 1997.

\bibitem{JAROSZKIEWICZ-97B}  Jaroszkiewicz G and Norton K, \emph{Principles
of discrete time mechanics: II. Classical field theory,} J. Phys. A: Math.
Gen., 30(7): 3145--3163, May 1997.

\bibitem{JAROSZKIEWICZ-99}  George Jaroszkiewicz, \emph{Discrete spacetime:
classical causality, prediction, retrodiction and the mathematical arrow of
time}, in V.~Di~Gesu, R.~Buccheri and M.~Saniga, editors, \emph{First
International Interdisciplinary Workshop: Studies on the Structure of Time:
from Physics to Psycho}(\emph{Patho})\emph{Logy}, 23-24 November 1999,
CNR-Area
della Ricerca di Palermo, Sicily, Kluwer Academic (New York), and in
gr-qc/0004026.

\bibitem{EPR}  B.~Podolsky A.~Einstein and N.~Rosen, \emph{Can quantum
mechanical description of reality be considered complete?} Phys. Rev.,
47: 777, 1935.

\bibitem{JAROSZKIEWICZ-97A}  Jaroszkiewicz G and Norton K, \emph{Principles
of discrete time mechanics: I. Particle systems,} J. Phys. A: Math. Gen.,
9(7): 3115--3144, May 1997.

\bibitem{DIRAC:64}  Dirac P A M, \emph{Lectures on Quantum Mechanics},
Yeshiva University (New York), Belfer Graduate School of Science Monograph
Series, no 2, 1964.
\end{thebibliography}
\end{document}